\documentclass[%
 aip,
 jmp,%
 amsmath,amssymb,
 reprint,%
]{revtex4-1}

\usepackage{graphicx}
\usepackage{dcolumn}
\usepackage{bm}

\begin{document}

\preprint{AIP/123-QED}

\title{Impact of degree heterogeneity on the behavior of trapping in Koch networks}

\author{Zhongzhi Zhang}
\email{zhangzz@fudan.edu.cn}

\author{Shuyang Gao}

\author{Wenlei Xie}


\affiliation {School of Computer Science, Fudan University, Shanghai
200433, China}

\affiliation {Shanghai Key Lab of Intelligent Information
Processing, Fudan University, Shanghai 200433, China}

\date{\today}

\begin{abstract}
Previous work shows that the mean first-passage time (MFPT) for
random walks to a given hub node (node with maximum degree) in
uncorrelated random scale-free networks is closely related to the
exponent $\gamma$ of power-law degree distribution $P(k)\sim
k^{-\gamma}$, which describes the extent of heterogeneity of
scale-free network structure. However, extensive empirical research
indicates that real networked systems also display ubiquitous degree
correlations. In this paper, we address the trapping issue on the
Koch networks, which is a special random walk with one trap fixed at
a hub node. The Koch networks are power-law with the characteristic
exponent $\gamma$ in the range between 2 and 3, they are either
assortative or disassortative. We calculate exactly the MFPT that is
the average of first-passage time from all other nodes to the trap.
The obtained explicit solution shows that in large networks the MFPT
varies lineally with node number $N$, which is obviously independent
of $\gamma$ and is sharp contrast to the scaling behavior of MFPT
observed for uncorrelated random scale-free networks, where $\gamma$
influences qualitatively the MFPT of trapping problem.
\end{abstract}

\maketitle

\begin{quotation}
As a fundamental dynamical process, random walks have received
considerable interest from the scientific community. Recent work
shows that the key quantity---mean first-passage time (MFPT) for
random walks to a given hub node (node with highest degree) on
uncorrelated random scale-free networks is qualitatively reliant on
the heterogeneity of network structure. However, in addition to the
power-law behavior, most real systems are also characterized by
degree correlations. In this paper, we study random walks on a
family of recently proposed networks---Koch networks that are
transformed from the well-known Koch curves and have simultaneously
power-law degree distribution and degree correlations with the power
exponent of degree distribution lying between 2 and 3. We explicitly
determine the MFPT, i.e., the average of first-passage time to a
target hub node averaged over all possible starting positions, and
show that the MFPT varies linearly with node number, independent of
the inhomogeneity of network structure. Our result indicates that
the heterogeneous structure of Koch networks has little impact on
the scaling of MFPT in the network family, which is in contrast with
result of MFPT previously reported for uncorrelated stochastic
scale-free graphs.
\end{quotation}

\section{Introduction}

In the past decade, a lot of endeavors have been devoted to
characterize the structure of real systems from the view point of
complex networks~\cite{AlBa02,DoMe02,Ne03,BoLaMoChHw06}, where nodes
represent system elements and edges interactions or relations
between them. One of the most important findings of extensive
empirical studies is that a wide variety of real networked systems
exhibit scale-free behavior~\cite{BaAl99}, characterized by a
power-law degree distribution $P(k) \sim k^{-\gamma}$ with degree
exponent $\gamma$ lying in the interval of $[2,3]$. Networks with
such broad tail distribution are called scale-free networks, which
display inhomogeneous structure encoded in the exponent $\gamma$:
the less the exponent $\gamma$, the stronger the inhomogeneity of
the network structure, and vice versa. The heterogeneous structure
critically influences many other topological properties. For
instance, it has been shown that in uncorrelated random scale-free
networks with node number $N$ (often called network order), their
average path length (APL)~\cite{WaSt98} relies on
$\gamma$~\cite{CoHa03,ChLu02}: for $\gamma=3$, $d(N) \sim \ln N$;
while for $2 \leq \gamma <3$, $d(N) \sim \ln \ln N$.

The power-law degree distribution also radically affects the
dynamical processes running on scale-free networks~\cite{DoGoMe08},
such as disease spreading~\cite{PaVe01},
percolation~\cite{CaNeStWa00}, and so on. Amongst various dynamics,
random walks are an important one that have a wide range of
applications~\cite{We1994,RoBe08,MeGolaMo08} and have received
considerable attention~\cite{Ga04,CoBeTeVoKl07,CoTr07,GoLa08}.
Recently, mean first-passage time (MFPT) for random walks to a given
target point in graphs, averaged over all source points has been
extensively
studied~\cite{Mo69,KaBa02PRE,KaBa02IJBC,Ag08,ZhQiZhGaGu10}. A
striking finding is that MFPT to a hub node (node with highest
degree) in scale-free networks scales sublinearly with network
order~\cite{ZhQiZhXiGu09,ZhGuXiQiZh09,AgBu09,TeBeVo09}, the root of
which is assumed to be the structure heterogeneity of the networks.
In particularly, it has been reported~\cite{KiCaHaAr08} that in
uncorrelated random scale-free networks, the MFPT $F(N)$ scales with
network order $N$ as $F(N) \sim N^{\frac{\gamma -2}{\gamma -1}}$.
However, real networks exhibit ubiquitous degree correlations among
nodes, they are either assortative or
disassortative~\cite{Newman02}. Then, an interesting question arises
naturally: whether the relation governing MFPT and degree exponent
$\gamma$ in uncorrelated scale-free networks is also valid for their
correlated counterparts.

In this paper, we study analytically random walks in the Koch
networks~\cite{ZhZhXiChLiGu09, ZhGaChZhZhGu10} that are controlled
by a positive-integer parameter $m$. This family of networks is
scale-free with the degree exponent $\gamma$ lying between 2 and 3,
and it may be either disassortative  ($m>1$) or uncorrelated
($m=1$). We focus on the trapping problem, a particular case of
random walks with a fixed trap located at a hub node. We derive
exactly the MFPT that is the average of first-passage time (FPT)
from all starting nodes to the trap. The obtained explicit formula
displays that in large networks with $N$ nodes, the MFPT grows
linearly with $N$, which is independent of $\gamma$ and, showing
that the structure inhomogeneity has no quantitative influence on
the MFPT to the hub in Koch networks, which lies in their symmetric
structure and other special features and is quite different from the
result previously reported for uncorrelated random scale-free
networks. Our work deepens the understanding of random walks
occurring on scale-free networks.

\begin{figure}
\begin{center}
\includegraphics[width=0.80\linewidth,trim=100 0 100 0]{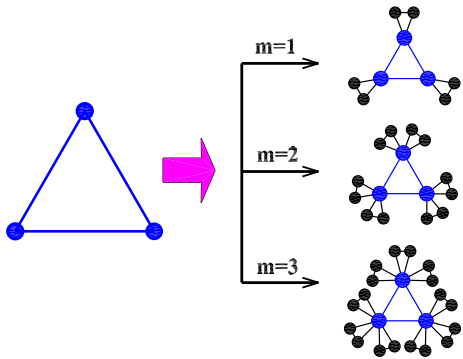}
\end{center}
\caption[kurzform]{\label{iterative} (Color online) Iterative
construction method for the Koch networks. }
\end{figure}

\section{Construction and properties of Koch networks}

\begin{figure}
\begin{center}
\includegraphics[width=1.0\linewidth,trim=100 0 100 0]{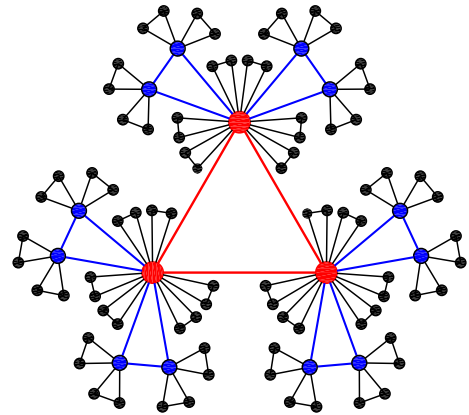}
\end{center}
\caption[kurzform]{\label{network2}(Color online) A network
corresponding to the case of $m=2$.}
\end{figure}

The Koch networks governed by a parameter $m$ are derived from the
famous Koch curves~\cite{Ko1906,LaVaMeVa87}, which are constructed
in an iterative way~\cite{ZhZhXiChLiGu09,ZhGaChZhZhGu10}. Let
$K_{m,t}$ denote the Koch networks after $t$ iterations. Then, the
networks can be generated as follows: Initially ($t=0$), $K_{m,0}$
is a triangle. For $t\geq 1$, $K_{m,t}$ is obtained from $K_{m,t-1}$
by adding $m$ groups of nodes for each of the three nodes of every
existing triangles in $K_{m,t-1}$. Each node group consists of two
nodes, both of which and their ``father'' node are connected to one
another shaping a new triangle. That is to say, to get $K_{m,t}$
from $K_{m,t-1}$, one can replace each existing triangle in
$K_{m,t-1}$ by the connected clusters on the right-hand side of
Fig.~\ref{iterative}. Figure~\ref{network2} show a network
corresponding to $m=2$ after several iterations.

By construction, the total number of triangles $L_\triangle(t)$ at
iteration $t$ is $L_\triangle(t)=(3m+1)^t$, and the number of nodes
created at iteration $t$ is
$L_v(t)=6m\,L_\triangle(t-1)=6m\,(3m+1)^{t-1}$. Then, the total
number of nodes $N_t$ present at step $t$ is
\begin{equation}\label{Nt}
N_t=\sum_{t_i=0}^{t}L_v(t_i)=2\,(3m+1)^{t}+1\,.
\end{equation}

Let $k_i(t)$ be the degree of a node $i$ at time $t$, which is added
to the networks at iteration (step) $t_i$ ($t_i\geq 0$). Then,
$k_i(t_i)=2$. Let $L_\triangle(i,t)$ denote the number of triangles
involving node $i$ at step $t$. According to the construction
algorithm, each triangle involving node $i$ at a given step will
give birth to $m$ new triangles passing by node $i$ at next step.
Thus, $L_\triangle(i,t)=(m+1)\,L_\triangle(i,t-1)=(m+1)^{t-t_{i}}$.
Moreover, it is easy to have $k_i(t)=2\,L_\triangle(i,t)$, i.e.,
\begin{equation}\label{ki}
k_i(t)=2\,L_\triangle(i,t)=2(m+1)^{t-t_{i}}\,,
\end{equation}
which implies
\begin{equation}\label{ki2}
k_i(t)=(m+1)\,k_i(t-1).
\end{equation}

The Koch networks present some common features of real
systems~\cite{AlBa02,DoMe02}. They are scale-free, having a
power-law degree distribution $P(k)\sim k^{-\gamma}$ with
$\gamma=1+\frac{\ln(3m+1)}{\ln(m+1)}$ belonging to the range between
2 and 3. Thus, parameter $m$ controls the extent of heterogeneous
structure of Koch networks with larger $m$ corresponding to more
heterogeneous structure. They have small-world effect with a low APL
and a high clustering coefficient. In addition, their degree
correlations can be also determined. For $m=1$, they are completely
uncorrelated; while for other values of $m$, the Koch networks are
disassortative.

\section{\label{sec:standwalk}Random walks with a trap fixed on a hub node}

\begin{figure}
\begin{center}
\includegraphics[width=0.9\linewidth,trim=100 10 100 10]{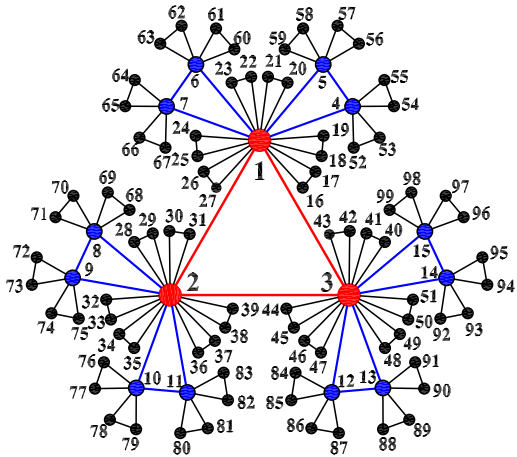}
\end{center}
\caption[kurzform]{\label{label} (Color online) labels of all nodes
in $K_{2,2}$.}
\end{figure}

After introducing the construction and structural properties of the
Koch networks, we continue to investigate random walks~\cite{NoRi04}
performing on them. Our aim is to uncover how topological features,
especially degree correlations influence the behavior of a simple
random walk on Koch networks with a single trap or a perfect
absorber stationed at a given node with highest degree. At each step
the walker, located on a given node, moves uniformly to any of its
nearest neighbors. To facilitate the description, we label all the
nodes in $K_{m,t}$ as follows. The initial three nodes in $K_{m,0}$
have label 1, 2, and 3, respectively. In each new generation, only
the newly created nodes are labeled, while all old nodes hold the
labels unchanged. That is to say, the new nodes are labeled
consecutively as $M+1, M+2,\ldots, M+\Delta M$, with $M$ being the
number of all pre-existing nodes and $\Delta M$ the number of newly
created nodes. Eventually, every node has a unique labeling: at time
$t$ all nodes are labeled continuously from 1 to $N_t=2(3m+1)^t+1$,
see Fig.~\ref{label}. We locate the trap at node 1, denoted by
$i_T$.

We will show that the particular selection of the trap location
makes it possible to compute analytically the relevant quantity of
the trapping process, i.e., mean first-passage time. Let $F_i^{(t)}$
denote the first-passage time of node $i$ in $K_{m,t}$ except the
trap $i_T$, which is the expected time for a walker starting from
$i$ to visit the trap for the first time. The mean of FPT
$F_i^{(t)}$ over all non-trap nodes in $K_{m,t}$ is MFPT, denoted by
$\langle F \rangle_t$, the determination of which is a main object
of the section. To this end, we first establish the scaling relation
governing the evolution of $F_i^{(t)}$ with generation $t$.

\begin{table}
\caption{First-passage time $F_i^{(t)}$ for a random walker starting
from node $i$ in $K_{m,t}$ for different $t$. Note that thanks to
the symmetry, nodes in the same column are equivalent to one
another, since they have the same FPT.} \label{tab:FPT1}
\begin{center}
\begin{tabular}{l|ccccccccc}
\hline \hline
            $t/i$ & $2-3$ & $4-7$ & $8-15$ & $16-27$ & $28-51$ & $52-67$ & $68-99$ \\
            \hline
            0 & $2$ \\
            1 & $14$ & $2$ & $16$ \\
            2 & $98$ & $14$ & $112$ & $2$ & $100$ & $16$ & $114$ \\
            3 & $686$ & $98$ & $784$ & $14$ & $700$ & $112$ & $798$ \\
            4 & $4802$ & $686$ & $5488$ & $98$ & $4900$ & $784$ & $5586$ \\
            5 & $33614$ & $4602$ & $38416$ & $686$ & $34300$ & $5488$ & $39102$ \\
         \hline \hline
\end{tabular}
\end{center}
\end{table}

\subsection{Evolution scaling for first-passage time}

We begin by recording the numerical values of $F_i^{(t)}$ for the
case of $m=2$. Clearly, for all $t \geq 0$, $F_1^{(0)}=0$; for
$t=0$, it is trivial, and we have $F_2^{(0)}=F_3^{(0)}=2$. For $t
\geq 1$, the values of $F_i^{(t)}$ can be obtained numerically but
exactly via computing the inversion of a matrix, which will be
discussed in the following text. Here we only give the values of
computation. In the generation $n =1$, by symmetry we have
$F_2^{(1)}=F_3^{(1)}=14$,
$F_4^{(1)}=F_5^{(1)}=F_6^{(1)}=F_7^{(1)}=2$, and
$F_8^{(1)}=F_9^{(1)}=\ldots=F_{15}^{(1)}=16$. Analogously, for
$t=2$, the numerical solutions are $F_2^{(2)}=F_3^{(2)}=98$,
$F_4^{(2)}=F_5^{(2)}=F_6^{(2)}=F_7^{(2)}=14$,
$F_8^{(2)}=F_9^{(2)}=\ldots=F_{15}^{(2)}=112$,
$F_{16}^{(2)}=F_{17}^{(2)}=\ldots=F_{27}^{(2)}=2$,
$F_{28}^{(2)}=F_{29}^{(2)}=\ldots=F_{51}^{(2)}=100$,
$F_{52}^{(2)}=F_{53}^{(2)}=\ldots=F_{67}^{(2)}=16$, and
$F_{68}^{(2)}=F_{69}^{(2)}=\ldots=F_{99}^{(2)}=114$.
Table~\ref{tab:FPT1} lists the numerical values of $F_i^{(t)}$ for
some nodes up to $t=5$.

The numerical values reported in Table~\ref{tab:FPT1} show that for
any node $i$, its FPT satisfies the relation
$F_i^{(t+1)}=(3m+1)\,F_i^{(t)}$. In other words, upon growth of Koch
networks from generation $t$ to $t+1$, the FPT of any node increases
to $3m+1$ times. For example,
$F_2^{(5)}=7\,F_2^{(4)}=7^2\,F_2^{(3)}=7^3\,F_2^{(2)}=7^4\,F_2^{(1)}=7^5\,F_2^{(0)}=33614$,
$F_8^{(5)}=7\,F_8^{(4)}=7^2\,F_8^{(3)}=7^3\,F_8^{(2)}=7^4\,F_8^{(1)}=38416$,
and so forth. This scaling is a basic property of random walks on
the family of Koch networks, which can be established based on the
following arguments.

\begin{figure}
\begin{center}
\includegraphics[width=.65\linewidth,trim=90 30 90 10]{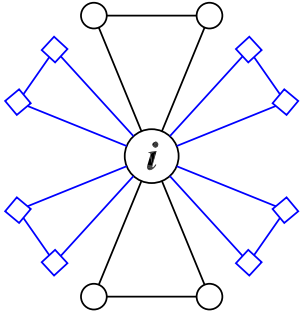}
\end{center}
\caption{(Color online) Growth of first-passage time in going from
$K_{m,t}$ to $K_{m,t+1}$ in the case of $m=2$. Node $i \in K_{m,t}$
has $k_i$ neighbor nodes in generation $t$ ($\bigcirc$) and $mk_i$
new neighbor nodes in generation $t+1$ ($\square$). A new neighbor
of node $i$ has a degree of 2, and is simultaneously linked to
another new neighbor of $i$.} \label{Evolution}
\end{figure}

Examine an arbitrary node $i$ in the Koch networks $K_{m,t}$.
Equation~(\ref{ki2}) shows that upon growth of the networks from
generation $t$ to $t+1$, the degree $k_i$ of node $i$ grows by $m$
times, i.e., it increases from $k_i$ to $(m+1)k_i$. Let $A$ denote
the FPT for going from node $i$ to any of its $k_i$ old neighbors,
and let $B$ be FPT for starting from any of the $mk_i$ new neighbors
of node $i$ to one of its $k_i$ old neighboring nodes. Then the
following equations can be established (see Fig.~\ref{Evolution}):
\begin{eqnarray}\label{FPT1}
\left\{
\begin{array}{ccc}
A&=&\frac{1}{m+1}+\frac{m}{m+1}(1+B),\\
B&=&\frac{1}{2}(1+A)+\frac{1}{2}(1+B),
 \end{array}
 \right.
\end{eqnarray}
which yield $A=3m+1$. This indicates when the networks grow from
generation $t$ to $t+1$, the FPT from any node $i$ ($i \in K_{m,t}$)
to any node $j$ ($j\in K_{m,t+1}$) increases on average $3m$ times.
Then, we have $F_i^{(t+1)}=(3m+1)\,F_i^{(t)}$ For explanation, see
Refs.~\cite{Bobe05,HaBe87} and related references therein. The
obtained relation for FPT is very useful for the following
derivation of MFPT.

\subsection{Explicit expression for mean first-passage time}

Having obtained the scaling dominating the evolution for FPT, we now
draw upon this relation to determine the MFPT, with an aim to derive
an explicit solution. For the sake of convenient description of
computation, we represent the set of nodes in $K_{m,t}$ as
$\Theta_t$, and denote the set of nodes created at generation $t$ by
$\bar{\Theta}_t$. Evidently, the relation $\Theta_t=\bar{\Theta}_t
\cup \Theta_{t-1}$ holds. In addition, for any $r \leq t$, we define
the two following variables:
\begin{equation}\label{MFPT01}
F_{r,\text{tot}}^{(t)}=\sum_{i \in \Theta_r} F_i^{(t)},
\end{equation}
and
\begin{equation}\label{MFPT02}
\bar{F}_{r,\text{tot}}^{(t)}=\sum_{i \in \bar{\Theta}_r} F_i^{(t)}.
\end{equation}
Then, we have
\begin{equation}\label{MFPT03}
F_{t,\text{tot}}^{(t)}=F_{t-1,\text{tot}}^{(t)}+\bar{F}_{t,\text{tot}}^{(t)}=(3m+1)F_{t-1,\text{tot}}^{(t-1)}+\bar{F}_{t,\text{tot}}^{(t)}\,,
\end{equation}
and
\begin{equation}\label{MFPT03}
\langle F \rangle_t=\frac{F_{t,\text{tot}}^{(t)}}{N_t-1}\,.
\end{equation}
Thus, to explicitly determine the quantity $\langle F \rangle_t$,
one should first find $F_{t,\text{tot}}^{(t)}$, which can be reduced
to determining $\bar{F}_{t,\text{tot}}^{(t)}$. Next, will show how
to solve the quantity $\bar{F}_{t,\text{tot}}^{(t)}$.

\begin{figure}[h]
\begin{center}
\includegraphics[width=.7\linewidth,trim=85 10 80 0]{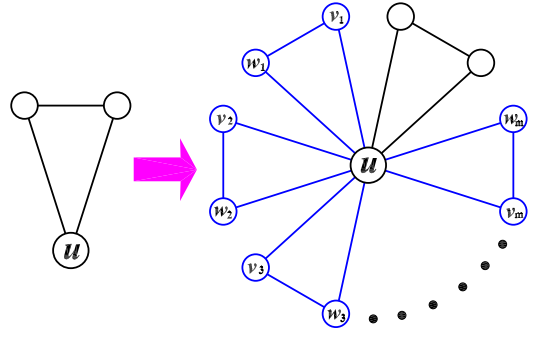}
\caption{(Color online) Illustration showing the relation of the
first-passage times for each pair of two new nodes ($v_x$ and $w_x$
with $x$=1,2,\ldots, or $m$) and the old node $u$ as one point of
the triangle generating the new nodes.}\label{iter}
\end{center}
\end{figure}

By construction, at a given generation, for each triangle passing by
node $u$, it will generate $m$ new triangles involving $u$ (see
Fig.~\ref{iter}). For each of the $m$ new triangles, the
first-passage times for its two new nodes ($v_x$ and $w_x$) and that
of its old node $u$ follow the relations:
\begin{eqnarray}\label{MFPT04}
\left\{
\begin{array}{ccc}
F(v_x) &=& 1+\frac{1}{2}\left[F(w_x)+F(u)\right]\,,\\
F(w_x) &=& 1+\frac{1}{2}\left[F(v_x)+F(u)\right]\,.
 \end{array}
 \right.
\end{eqnarray}
In Eq.~(\ref{MFPT04}), $F(s)$ represents the expected time of a
particle to first visit the trap node, given that it starts from
node $s$. Equation~(\ref{MFPT04}) yields
\begin{equation}\label{MFPT05}
F(v_x)+F(w_x) = 4 +2F(u)\,.
\end{equation}
Summing Eq.~(\ref{MFPT05}) over all the $L_\triangle(t)=(3m+1)^t$
old triangles pre-existing at the generation $t$ and the three old
nodes of each of the  $L_\triangle(t)$ triangles, we obtain
\begin{eqnarray}\label{MFPT06}
\bar{F}_{t+1, {\rm tot}}^{(t+1)} &=& 3 \cdot 4 \cdot m
L_\triangle(t) +
\sum_{i \in \Theta_{t}}\left(2mL_\triangle(i,t)\cdot F_i^{(t+1)}\right) \nonumber\\
&=&12m(3m+1)^t + 2m\bar{F}_{t, {\rm tot}}^{(t+1)}
+ 2m(m+1)\bar{F}_{t-1, {\rm tot}}^{(t+1)}\nonumber\\
&\quad&+\ldots+2m(m+1)^{t-1}\bar{F}_{1, {\rm
tot}}^{(t+1)}+2m(m+1)^{t}\bar{F}_{0,{\rm tot}}^{(t+1)}\,.\nonumber\\
\end{eqnarray}
For instance, in $K_{2,2}$ (see Fig.~\ref{label}), $\bar{F}_{2, {\rm
tot}}^{(2)}$ can be expressed as
\begin{equation}\label{MFPT07}
\bar{F}_{2, {\rm tot}}^{(2)} = \sum_{i=16}^{99} F_{i}^{(2)} = 1176 +
12\,\bar{F}_{1, {\rm tot}}^{(2)}+ 36\,\bar{F}_{0, {\rm
tot}}^{(2)}\,.
\end{equation}

Now, we can determine $\bar{F}_{t, {\rm tot}}^{(t)}$ through a
recurrence relation, which can be obtained easily. From
Eq.~(\ref{MFPT07}), it is not difficult to write out $\bar{F}_{t+2,
{\rm tot}}^{(t+2)}$ as
\begin{eqnarray}\label{MFPT08}
\bar{F}_{t+2, {\rm tot}}^{(t+2)} &=& 12m(3m+1)^{t+1} +
2m\bar{F}_{t+1,{\rm tot}}^{(t+2)}
+ 2m(m+1)\bar{F}_{t, {\rm tot}}^{(t+2)}\nonumber\\
&\quad&+\ldots+2m(m+1)^{t}\bar{F}_{1, {\rm
tot}}^{(t+2)}+2m(m+1)^{t+1}\bar{F}_{0,{\rm tot}}^{(t+2)}\,.\nonumber\\
\end{eqnarray}
Equation~(\ref{MFPT08}) minus Eq.~(\ref{MFPT06}) times $(m+1)(3m+1)$
and using the relation $F_i^{(t+2)}=(3m+1)\,F_i^{(t+1)}$, we have
\begin{equation}\label{MFPT09}
\bar{F}_{t+2,{\rm tot}}^{(t+2)}=(3m+1)^2\bar{F}_{t+1, {\rm
tot}}^{(t+1)}-12m^2(3m+1)^{t+1}\,.
\end{equation}
Making use of the initial condition $\bar{F}_{1,{\rm
tot}}^{(1)}=24m^2+20m$, Eq.~(\ref{MFPT09}) is solved inductively to
yield
\begin{equation}\label{MFPT10}
\bar{F}_{t,{\rm
tot}}^{(t)}=4m(3m+1)^{t-1}+(24m^2+16m)(3m+1)^{2t-2}\,.
\end{equation}

Inserting Eq.~(\ref{MFPT10}) for $\bar{F}_{t,{\rm tot}}^{(t)}$ into
Eq.~(\ref{MFPT03}), we have
\begin{eqnarray}\label{MFPT11}
F_{t,\text{tot}}^{(t)}&=&(3m+1)\,F_{t-1,\text{tot}}^{(t-1)}+\nonumber\\
&\quad&4m(3m+1)^{t-1}+(24m^2+16m)(3m+1)^{2t-2}\,.\nonumber\\
\end{eqnarray}
Since $F_{0,\text{tot}}^{(0)}=4$, we can resolve Eq.~(\ref{MFPT11})
by induction to obtain
\begin{equation}\label{MFPT12}
F_{t,\text{tot}}^{(t)}=\frac{4}{3}
(3m+1)^{t-1}[(6m+4)(3m+1)^t+3mt+3m-1]\,.
\end{equation}
By plugging Eq.~(\ref{MFPT12}) into Eq.~(\ref{MFPT03}), we obtain
the closed-form solution to the MFPT for random walks on the Koch
networks with an immobile trap stationed at a hub node:
\begin{equation}\label{MFPT13}
\langle F \rangle_t=\frac{2}{3(3m+1)}[(6m+4)(3m+1)^t+3mt+3m-1]\,.
\end{equation}

\subsection{Numerical calculations}

We have corroborated our analytical formula for MFPT provided by
Eq.~(\ref{MFPT13}) against direct numerical calculations via
inverting a matrix~\cite{KeSn76}. Indeed, the Koch network family
$K_{m,t}$ can be represented by its adjacency matrix $\textbf{A}_t$
of an order $N_t \times N_t$, the element $a_{ij}(t)$ of which is
either 1 or 0 defined as follows: $a_{ij}(t)=1$ if nodes $i$ and $j$
are directly connected by a link, and $a_{ij}(t)=0$ otherwise. Then
the degree, $d_{i}(t)$, of node $i$ in $K_{m,t}$ is given by
$d_{i}(t)=\sum_{j}^{N_t}a_{ij}(t)$, the diagonal degree matrix
$\textbf{Z}_t$ associated with $K_{m,t}$ is $\textbf{Z}_t={\rm diag}
(d_1(t), d_2(t),\ldots, d_i(t), \ldots, d_{N_t}(t))$, and the
normalized Laplacian matrix of  $K_{m,t}$ is provided by
$\textbf{L}_t=\textbf{I}_t-\textbf{Z}_t^{-1}\textbf{A}_t$, in which
$\textbf{I}_t$ is the $N_t \times N_t$ identity matrix.

Note that the random walks considered above is in fact a Markovian
process, and the fundamental matrix of Markov chain representing
such unbiased random walks is the inverse of a submatrix of
$\textbf{L}_t$, denoted by $\bar{\mathbf{L}}_t$ that is obtained by
removing the first row and column of $\textbf{L}_t$ corresponding to
the trap node. According to previous result~\cite{KeSn76}, the FPT
$F_i^{(t)}$ can be expressed by in terms of the entry
$\bar{l}_{ij}^{-1}(t)$ of $\bar{\mathbf{L}}_t^{-1}$ as
\begin{equation}\label{MFPT16}
F_i^{(t)}=\sum_{j=2}^{N_t}\bar{l}_{ij}^{-1}(t)\,,
\end{equation}
where $\bar{l}_{ij}^{-1}(t)$ is the expected times that the walk
visit node $j$, given that it starts from node $i$~\cite{KeSn76}.
Using Eq.~(\ref{MFPT16}) we can determine $F_i^{(t)}$ numerically
but exactly for different non-trap nodes at various generation $t$,
as listed in Table~\ref{tab:FPT1}.

\begin{figure}
\begin{center}
\includegraphics[width=0.85\linewidth, trim=65 90 150 65]{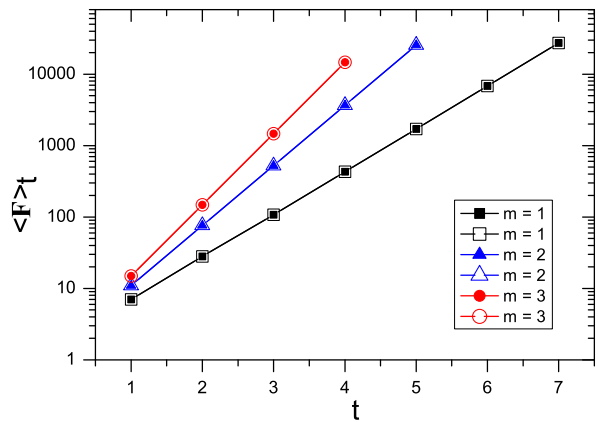}
\end{center}
\caption[kurzform]{\label{Time} (Color online) Mean first-passage
time $\langle F \rangle_t$ as a function of the generation $t$ on a
semilogarithmic scale for different values of $m$. The empty symbols
represent the numerical results obtained by direct calculation from
Eq.~(\ref{MFPT17}); while the filled symbols correspond to the
rigorous values provided by Eq.~(\ref{MFPT13}).}
\end{figure}

By definition, the MFPT $\langle F \rangle_t$ that is the mean of
$F_i^{(t)}$ over all initial non-trap nodes in $K_{m,t}$ reads as
\begin{eqnarray}\label{MFPT17}
\langle F \rangle_t&=&\frac{1}{N_t-1}\sum_{i=2}^{N_t}F_i^{(t)}\nonumber \\
&=&\frac{1}{2(3m+1)^{t}}\sum_{i=2}^{N_t}\sum_{j=2}^{N_t}\bar{l}_{ij}^{-1}(t)\,.
\end{eqnarray}
In Fig.~\ref{Time}, we compare the analytical results given by
Eq.~(\ref{MFPT13}) and the numerical results obtained by
Eq.~(\ref{MFPT17}) for various $t$ and $m$. Figure~\ref{Time} shows
that the analytical and numerical values for $\langle F \rangle_{t}$
are in full agreement with each other. This agreement serves as a
test of our analytical formula.

\subsection{Dependence of mean first-passage time on network order}

Below we will show how to express $\langle F \rangle_t$ as a
function of network order $N_t$, with the aim of obtaining the
relation between these two quantities. Recalling Eq.~(\ref{Nt}), we
have $(3m+1)^t=\frac{N_t-1}{2}$ and
$t=\frac{\ln(N_t-1)-\ln2}{\ln(3m+1)}$. Thus, Eq.~(\ref{MFPT13}) can
be recast as in terms of $N_t$ as
\begin{eqnarray}\label{MFPT14}
\langle F \rangle_t&=&
\frac{2(3m+2)}{3(3m+1)}(N_t-1)+\frac{2m[\ln(N_t-1)-\ln2]}{(3m+1)\ln(3m+1)}\nonumber\\
&\quad&\frac{2(3m-1)}{3(3m+1)}\,.
\end{eqnarray}
In the thermodynamic limit ($N_t \rightarrow \infty$), we have
\begin{equation}\label{MFPT15}
\langle F \rangle_t \approx \frac{2(3m+2)}{3(3m+1)}(N_t-1) \sim
N_t\,,
\end{equation}
showing that the MFPT grows linearly with increasing order of the
Koch networks. Equations~(\ref{MFPT14}) and~(\ref{MFPT15}) imply
that although for different $m$ the MFPT of whole family of Koch
networks is quantitatively different, it exhibits the same scaling
behavior despite the distinct extent of structure inhomogeneity of
the networks, which may be attributed to the symmetry and particular
properties of the networks studied.

It is known that the exponent $\gamma$ characterizing the
inhomogeneity of networks affects qualitatively the scaling of MFPT
for diffusion in random uncorrelated scale-free
networks~\cite{KiCaHaAr08}. Concretely, in random uncorrelated
scale-free networks with large order $N$, the MFPT $F(N)$ grows
sublinearly or linearly with network order as $F(N) \sim
N^{\frac{\gamma-2}{\gamma-1}}$ for all $\gamma >2$, which strongly
depends on $\gamma$. However, as shown above, in the whole family of
Koch networks, the MFPT displays a linear dependence on network
order, which is independent of $\gamma$, showing that the
inhomogeneity of structure has no quantitative impact on the scaling
behavior of MFPT for trapping process in Koch networks. Our obtained
result means that the scaling observed for MFPT in the
literature~\cite{KiCaHaAr08} is not a generic feature of all
scale-free networks, at least it is not valid for the Koch networks,
even for the case of $m=1$ when network is uncorrelated.

\section{Conclusions}

Power-law degree distribution and degree correlations play a
significant role in the collective dynamical behaviors on scale-free
networks. In this paper, we have investigated the trapping issue,
concentrating on a particular case with the trap fixed on a node
with highest degree on the Koch networks that display synchronously
a heavy-tailed degree distribution with general exponent $\gamma \in
[2,3]$ and degree correlations. We obtained explicitly the formula
for MFPT to the trapping node, which scales lineally with network
order, independent of the exponent $\gamma$. Our result shows that
structural inhomogeneity of the Koch networks has no essential
effect on the scaling of MFPT for the trapping issue, which departs
a little from that one expects and is as compared with the scaling
behavior reported for stochastic uncorrelated scale-free networks.
Thus, caution must be taken when making a general statement about
the dependence of MFPT for trapping issue on the inhomogeneous
structure of scale-free networks. Finally, it should be also
mentioned that both random uncorrelated networks and the Koch
networks addressed here cannot well describe real systems, future
work should focus on trapping problem on those networks better
mimicking realties. Anyway, our work provides some insight to better
understand the trapping process in scale-free graphs.

\begin{acknowledgments}
This work was supported by the National Natural Science Foundation
of China under Grants Nos. 60704044 and 61074119, and the Shanghai
Leading Academic Discipline Project No.B114. S. Y. G. also
acknowledges the support by Fudan's Undergraduate Research
Opportunities Program.
\end{acknowledgments}

\nocite{*}
\bibliography{aipsamp}

\end{document}